# Steering optical comb frequency by rotating polarization state

Y. Zhang[1,2], L. Yan[1], X. F. Zhang[1], L. Zhang[1], W. Han[1], W. Guo[1,3], H. Jiang[1,*] and S. Zhang[1]

1. Key Laboratory of time & frequency primary standards, National time service center, Xi'an 710600, China
2. University of the Chinese academy of sciences, Beijing 100049, China
3. School of science, Xi'an Shiyou University, Xi'an 710065, China
*email: haifeng.jiang@ntsc.ac.cn

**Optical frequency combs, with precise control of repetition rate and carrier-envelope-offset frequency, have revolutionized many fields, such as fine optical spectroscopy[1], optical frequency standards[2], ultra-fast science research[3], ultra-stable microwave generation[4] and precise ranging measurement[5]. However, existing high bandwidth frequency control methods have small dynamic range, requiring complex hybrid control techniques[6-9]. To overcome this limitation, we develop a new approach, where a home-made intra-cavity electro-optic modulator tunes polarization state of laser signal rather than only optical length of the cavity, to steer frequencies of a nonlinear-polarization-rotation[10] mode-locked laser[11]. By taking advantage of birefringence of the whole cavity, this approach results in not only broadband but also relative large-dynamic frequency control. Experimental results show that frequency control dynamic range increase at least one order in comparison with the traditional intra-cavity electro-optic modulator technique. In additional, this technique exhibits less side-effect than traditional frequency control methods.**

High bandwidth control is the thumb rule to have tight frequency-stabilization. The repetition rate ($f_r$) is generally controlled by tuning cavity length using a Piezotransducor (PZT) with a low bandwidth of around 1 kHz[12]. To suppress high frequency noise (up to 1 MHz), $f_r$ can be further controlled by tuning optical length of the cavity with an intra-cavity electro-optic modulator (EOM)[6]. However, the intra-cavity EOM has to be long enough to achieve sufficient control dynamic (Hertz scale)[13,14]; this significantly changes the cavity parameters, resulting in difficulties of mode-locking. On the other hand, the carrier-envelope-offset frequency ($f_{ceo}$) is typically controlled by tuning pumping power[15,16] with a relatively low bandwidth (~100 kHz typically) limited by gain lifetime. Techniques beyond this limit, such as extra-cavity frequency shift[7] and intra-cavity power-modulation[8], are rear to be adopted due to their small dynamic range of below a few MHz and 1 Hz, respectively.

We propose a method of rotating polarization state to steer frequency by using a special home-made intra-cavity EOM. In a fiber-based nonlinear-polarization-rotation (NPR) system, a combination of NPR due to Kerr effect and polarization state selection[17] leads to mode-locking operation within a certain range of polarization state setup. This range corresponds to a control dynamic range of comb frequencies. Due to the effects of fiber's birefringence, a change of power ratio between slow and fast axis, together with polarization state changes, leads to a group-velocity shift. On the other hand, the phase velocity, dominated by gain and nonlinear effects, should be relatively stable in the case of deep saturation condition. Consequently, $f_{ceo}$, representing difference between group velocity and phase velocity, is also sensitive to polarization state. This approach not only has broadband frequency control ability like what traditional intra-cavity EOM technique does, but also significantly enlarges the control dynamic range. To maximum this effect, polarization state shift should occur at the position right after the component for polarization state selection.

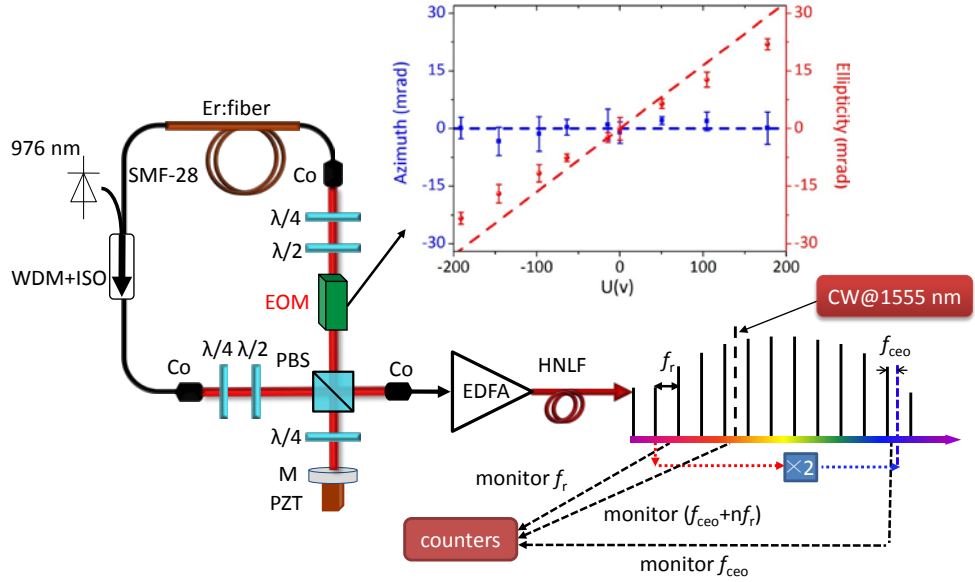

**Figure 1 Experimental setup.** The Er:fiber femtosecond laser source has a ring cavity including four wave plates, a polarization beam splitter (PBS), three collimators (CO), a wavelength division multiplexer (WDM) with an isolator (ISO), an electro-optic modulator (EOM) and fibers. A PZT mounted on a mirror (M) is employed to adjust the cavity length. A high nonlinear fiber (HNLF) and an erbium-doped fiber amplifier (EDFA) are used to produce the $f_{ceo}$ signal. Three frequency counters record the shifts of $f_r$, $f_{ceo}$, and a comb teeth at 1550 nm, respectively. Here, we want to note that the EOM (3-mm thick and 3×5 mm$^2$ LiNbO$_3$ crystal) is inserted right after the PBS in the free space optical path to rotate the polarization state of the laser signal. The inset shows polarization state shift in ellipticity (theoretical data: red dashed line, experimental data: red points) and azimuth (theoretical data: blue dashed line, experimental data: blue points) as a function of voltage on the EOM. Details on the mode-locked laser is in Methods part.

Figure 1 shows the experimental setup, where the ring laser has a repetition rate of about 192 MHz. The PBS is the component for polarization state selection and output coupling. The $f_r$, $f_{ceo}$ and the frequency of the mode nearby a reference laser (1550 nm) of this femtosecond laser are monitored by using frequency counters. A home-made 3-mm thick EOM right after the PBS is used as the polarization state controller. Different from typical EOM, this one (see Methods part) tunes the polarization state with a coefficient of ~ 0.12 mrad/v, which is measured in the range from -200 v to 200 v.

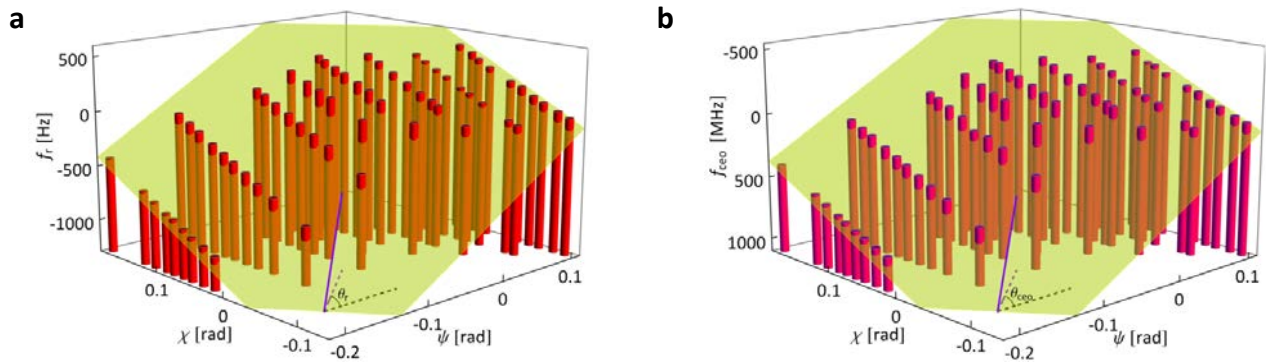

**Figure 2 Relation between comb frequencies and polarization state.** (a) the variation of repetition rate $f_r$ as a function of ellipticity χ and azimuth ψ ; (b) the variation of carrier-envelope-offset frequency $f_{ceo}$ as a function of ellipticity χ and azimuth ψ;

For (a) and (b), the maximum directions, denoted by $\theta_r$ and $\theta_{ceo}$, are 0.68 rad and 0.67 rad from azimuth axis, so that one can control the frequencies with maximum sensitivities of ~ 6 kHz/rad and -5000 MHz/rad, respectively.

Figure 2 shows the measured data of the relation between comb frequencies and polarization, indicating that such frequencies are sensitive to both ellipticity $\chi$ and azimuth $\psi$. In the experiments, we tune the ellipticity $\chi$ and azimuth $\psi$ in a range of 0.24 rad and 0.28 rad respectively by acting bias voltage on the EOM and rotating the waveplates following the EOM. Beyond this range, the signal-to-noise ratio of $f_{ceo}$ starts to drop down and/or a continuous-wave (CW) signal appears in optical spectrum. The dynamic ranges of $f_r$ and $f_{ceo}$ are about 1 kHz and 850 MHz in this range. These ranges are comparable to that of commonly-used large dynamic frequency control techniques (i.e. $f_r$ control with a PZT and $f_{ceo}$ control by tuning pumping power), implying that it is possible to get rid of complex hybrid methods for comb frequencies steering.

For Fig. 2 (a), all the measured points distribute in a tilt plane, where sensitivities of $f_r$ are ~3 kHz/rad in ellipticity and ~5 kHz/rad in azimuth, exhibiting a maximum sensitivity of ~6 kHz/rad. The birefringence level of the ring cavity, dominated by erbium-doped fiber, is experimentally estimated to be ~320 fs. It can induce a relative repetition rate shift of ~ 6.2E-5, corresponding to a maximum sensitivity of 12 kHz/rad. The reduction factor of 0.5 is attributed to the reason that the polarization state rotation is not along with the maximum efficient direction. In this case, nonlinear effects (like Kerr effect) play a less important role. While the polarization is tuned 0.05 rad in ellipticity, the laser`s output increases from 144 mW to144.2 mW. Such a 0.2 mW power change can be induced by 1 mW pumping power variation. However, such a pumping power change shifts $f_r$ and $f_{ceo}$ only about -12 Hz and 0.9 MHz, which are at least one order of magnitude below polarization shift effect. The $f_r$ increases with the output power, indicating more power transfer along the fast axis. It is consistent with the previous theoretical results that the intra-cavity laser signal pass along the slow axis of the ring cavity[18]. For Fig.2 (b), the sensitivities of $f_{ceo}$ are about -2700 MHz/rad in ellipticity and about -4200 MHz/rad in azimuth, exhibiting a maximum sensitivity of about -5000 MHz/rad. It is clear that tilt plates in both sub-figures are very similar, giving that the comb tooth at 1.5 μm (i.e. $f_{ceo}+nf_r$) is relatively stable, which is in agreement with our previous assumption. Actually, we also find that the peak wavelength is almost unchanged during the experiment, which supports the assumption as well.

Now, we can safely conclude that the polarization, including the ellipticity $\chi$ and azimuth $\psi$, can be regarded as two "degrees of freedom", which can be used to steer the optical comb frequencies with both broad bandwidth and relative large dynamic range.

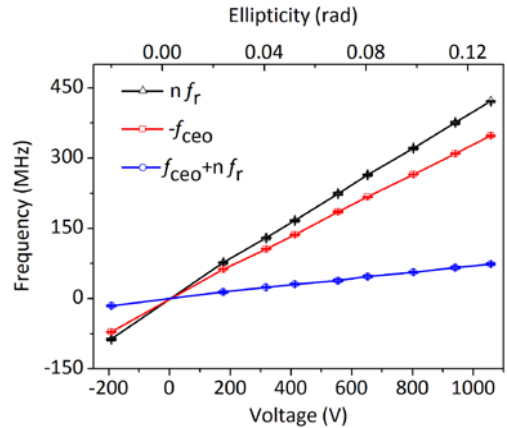

**Figure 3 Variation of comb frequencies versus voltage acting on the intra-cavity EOM.** Frequency variations of a comb teeth @ 1550 nm ($nf_r$ ,+$f_{ceo}$), $f_{ceo}$ and $nf_r$ are shown with blue, red, and black lines, respectively, where n is about $1\times 10^6$.

As frequency control is realized by using EOM, we measure the response of comb frequencies to voltage applied on the EOM, which changes polarization state only in ellipticity. Fig. 3 shows the variation of a comb teeth @ 1550 nm ($nf_r$ ,+$f_{ceo}$), $f_{ceo}$ and $nf_r$ as a function of voltage, giving corresponding coefficients to be about 0.07 MHz/v, -0.33 MHz/v and 0.40 MHz/v. Here we want to note that control coefficients of $f_r$, $f_{ceo}$, contrary to the traditional intra-cavity EOM technique[19] with the same size of EOM, are ~7 and ~50 times larger, although this case does not present the maximum efficiency of frequency control. Furthermore, we observe that frequency of the comb teeth is relatively stable, resulting in a small cross talk between control servo loops of $f_{ceo}$ and the tooth's frequency. Another advantage of this frequency control approach is that the power fluctuation introduced by $f_{ceo}$ control is only

1/270th of the pumping power control method. This is good for frequency combs utilization because some applications are sensitive to power variation. All above indicates that this technique is very suitable for $f_{ceo}$ control.

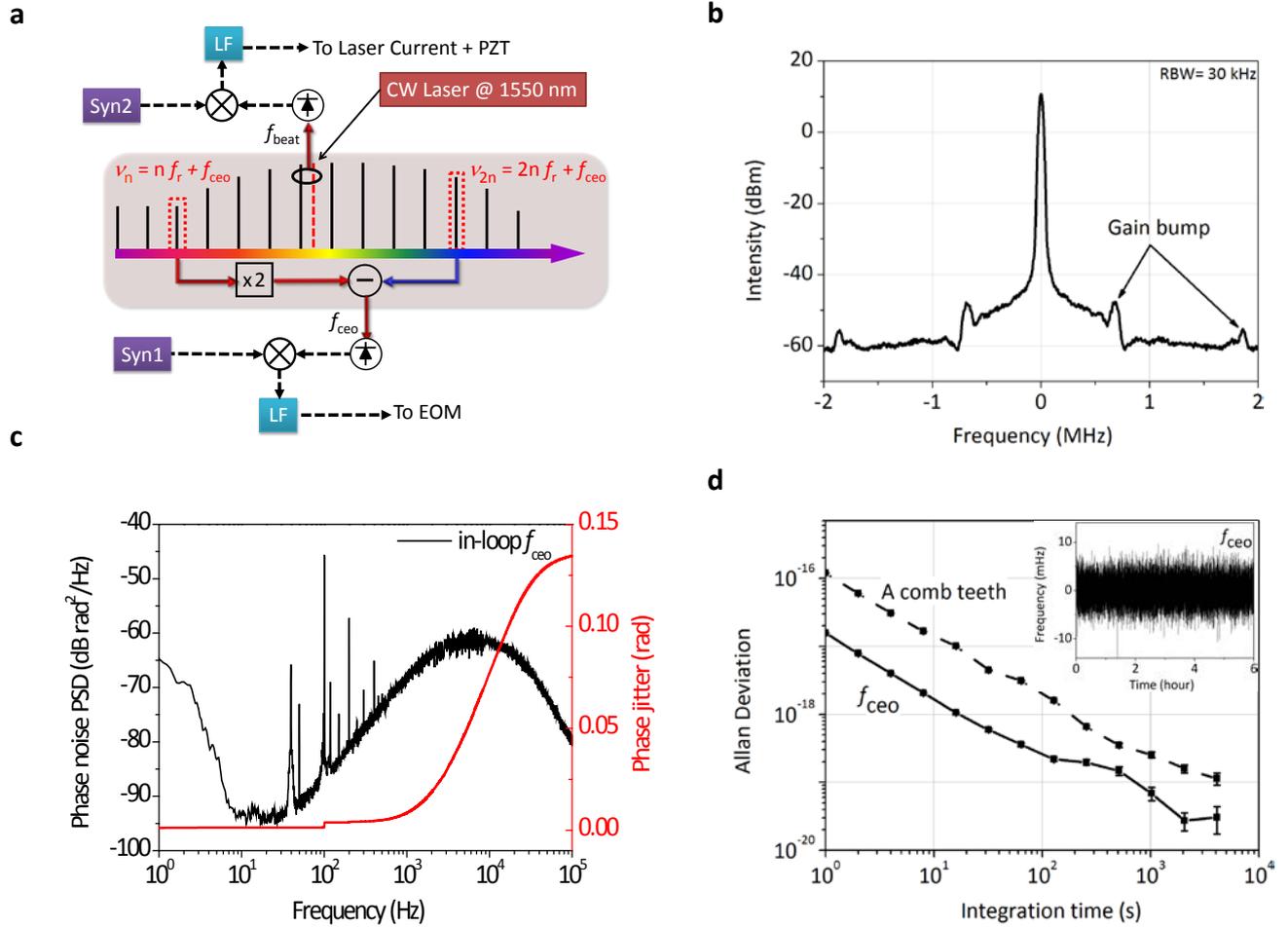

**Figure 4 setup and results of comb frequency stabilization with the new approach.** a, Schematic of comb's frequency stabilization, where $f_{ceo}$ is phase-locked onto a synthesizer (Syn1) with the new technique, and a comb teeth (~193 THz) is phase-locked onto a continuous-wave (CW) laser (1550 nm) with a frequency shift produced by Syn2. Loop filters (LF) are proportional-integral amplifiers. b, spectrum of in-loop $f_{ceo}$ while it is phase-stabilized. c, phase noise power spectral density (PSD) and phase jitter of in-loop phase-stabilized $f_{ceo}$. d, in-loop relative frequency instability of $f_{ceo}$ (solid black square) and a comb teeth (dashed black square). Inset is raw data of $f_{ceo}$ fluctuation recorded with frequency counter. Note that the used frequency counters are Π–type ones made by K&K company under averaging-mode.

To verify frequency steering ability of our approach, we stabilize $f_{ceo}$ onto a RF reference signal with the EOM, and a comb teeth onto a CW laser at 1550 nm simultaneously by controlling pumping power and the PZT, as shown in Fig. 4. Gain bumps show up at ~650 kHz and ~1.8 MHz, indicating that $f_{ceo}$ control loop has a broad servo bandwidth. Phase jitter from 1 Hz to 100 kHz is about 0.13 rad. Peak in-loop frequency jitter, recorded with a frequency counter, is well below 10 mHz at 1 second gate time; frequency stability normalized by 193 THz (1550 nm) is about $2\times10^{-17}$ @ 1 s and scales down with a slope of $1/\tau$ for short terms. Relative frequency stability of $f_r$ is about 10 time higher due to the response speed of pumping power control is relatively slow. Cross talk effect between two frequency control loops cannot be observed even under high gain conditions, while it was the dominating limitation of our previous system with the traditional intra-cavity EOM technique[19].

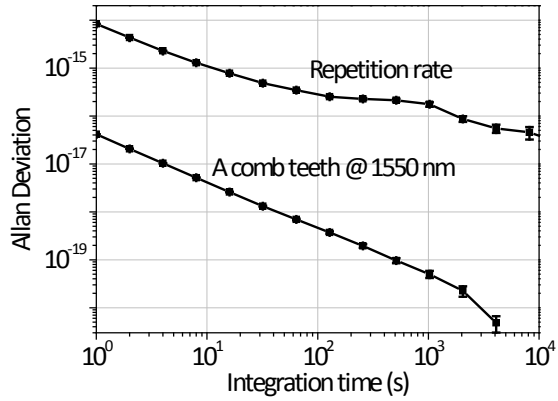

**Figure 5 in-loop relative frequency instability of a comb teeth and the repetition rate.** the comb teeth (~193 THz) is phase-locked onto a CW laser with the EOM; while the repetition rate is phase-locked onto a 9.2 GHz frequency reference.

This technique can also be used for frequency stabilization of $f_r$ or a comb teeth. As shown in Fig. 5, in-loop relative frequency instability of the comb teeth is $2\times10^{-17}$ @ 1 s; while that of $f_r$ is $8\times10^{-15}$ @ 1 s, limited by noise floor of the measurement system.

In summary, we have developed a new technique of optical comb frequencies control by tuning the polarization state of the laser. This approach can steer frequencies in broad bandwidth and a large dynamic range by taking advantage of birefringence of the whole laser cavity. In addition, it exhibits less side effects such as less cross talk and less power variation during frequency control. Our system does not demonstrate the maximum frequency turning ability due to the home-made EOM rotates polarization only in ellipticity. Moreover, improvement for frequency control coefficient could be easily done by employing high birefringence fibers. With this technique, hybrid frequency control is no more mandatory for tight comb frequency stabilization. It is worth noting that polarization is a new freedom for comb frequency control, and this technique works well in a NPR-based comb system, whether or not it can be applied to other types combs is an open question.

**Methods:**

**The Er:fibre-based femtosecond laser.** The home-made mode-locked laser, pumped with two 980 nm 500 mW pigtailed diode lasers, has a repetition rate of 192 MHz. The output power is ~ 140 mW. Peak power wavelength is at 1575 nm, and full width half maximum of the spectrum is ~ 20 nm. The ring cavity of the laser, as shown in figure 1, includes three types of fibers: 41 cm (Er110-4/125) erbium-doped fiber, 48 cm SMF-28 fiber and 5 cm HI1060 fiber. Net dispersion of the laser cavity is estimated to be slightly negative (~-5000 fs$^2$). $f_{ceo}$, produced with a common pass $f$-$2f$ interferometer, has a signal-to-noise ratio of ~ 40 dB (300 kHz resolution). The PZT (0.5 cm) driven by a tunable high voltage signal generator (0 v, +150 v) tunes the reflection mirror in a range of ~5 μm, corresponding to a repetition rate shift of ~1 kHz. The EOM is driven with a high voltage signal generator (-200 v, +200 v) with modulation port for response measurement and stabilization of the comb frequencies. In the experiment, another high voltage signal generator (0, +1000 v) without modulation function is used for frequency control dynamic range evaluation.

**EOM design.** The EOM is made of LiNbO$_3$ crystal which is single optical axis material in the absence of an applied electric field for the crystal. The direction of the laser passing is the crystal optical axis (defined as z-axis). The crystal's size is 3(x)×5(y) mm$^2$ and 3(z)-mm thick. The external electric field is parallel to x axis, and two yz surface are coated with gold. Incidence light has a linear polarization state alinement to x-axis. We simulate polarization rotation induced by external electric field, based on the refractive index ellipsoid theory[20]. The phase change between x-axis and y-axis of the incident light is given by $\Delta\varphi= (2\pi/\lambda)Lr_{22}Vn_o^3/d$, where $r_{22}$ is the nonzero component of the electro-optic tensor for LiNbO$_3$, $n_o$ is the index of refraction of LiNbO$_3$ crystal, $V$ is the voltage applied across the EOM, and $d$ is the distance between the electrodes on the EOM. Then using the Jones matrix and stokes vectors[21,22], we deduce that the EOM's application to polarization rotation in azimuth and in ellipticity can be expressed as $\tan2\psi = \tan2\theta/\cos\varphi$ and $\sin2\chi=\cos2\theta\sin\varphi$, respectively. $\psi$ is the azimuth angle, $\chi$ is the ellipticity angle, and $\theta$ is the angle between the initial polarization direction and x axis. Here, the polarization of the incident light is perpendicular to x axis ($\theta=0$). As shown in fig. 1, by tuning bias voltage on the EOM, the theoretical polarization rotation in ellipticity and in azimuth are 0.14 mrad/v and null respectively, in agreement with the experimental result of 0.12 mrad/v.

**Birefringence of the ring cavity.** Coefficients of fibers' birefringence do not exist in the datasheet. To simplify estimation process of the laser cavity`s birefringence level, we suppose that all birefringence is attributed to Er110 fiber due to the following two reasons. First of all, polarization mode dispersion of the SMF-28 fiber is given in the datasheet to be 0.1 ps/km$^{1/2}$, and coherent length of polarization stage is tens meters[23]. Thus, we can estimate that a short (< 1 m) SMF-28 having a birefringence coefficient of 1-3 fs/m. In this case, the 0.48 cm SMF-28 fiber introduces a birefringence of 0.5 – 1.5 fs, which is negligible. Secondly, HI1060 fiber is short and should have smaller birefringence coefficient than that of the Er110 fiber. To evaluate the birefringence level of Er110 fiber, we develop a figure-8 mode-locked laser with a nonlinear amplifier loop mirror (NALM) comprising 41 cm highly erbium-doped fiber, 19.5 cm SMF-28 fiber and 5 cm HI1060 fiber. Birefringence of the NALM causes spectral filtering effect, resulting in a serial of equivalent width fringes exhibiting in the optical spectrum[24]. The width of these fringes is about 25 nm, giving a birefringence of 320 fs from (1550nm)$^2$/25nm/3E8m/s. Thus, Er110 fiber has a birefringence coefficient of ~ 667 fs/m.